\documentclass[conference]{IEEEtran}
\IEEEoverridecommandlockouts
\usepackage{cite}
\usepackage{amsmath,amssymb,amsfonts}
\usepackage{amsthm}
\usepackage{multirow}
\usepackage{booktabs}
\usepackage{subfigure}
\usepackage{tikz}
\usetikzlibrary{fit,positioning}
\usepackage{float}
\usepackage[switch]{lineno}
\usepackage{algorithmic}
\usepackage{graphicx}
\usepackage{textcomp}
\usepackage{xcolor}
\def\BibTeX{{\rm B\kern-.05em{\sc i\kern-.025em b}\kern-.08em
    T\kern-.1667em\lower.7ex\hbox{E}\kern-.125emX}}
\begin{document}

\title{ARGO: Modeling Heterogeneity in E-commerce Recommendation\\
{\footnotesize }
\thanks{$^*$ Corresponding author.\par
The first three authors contribute equally to this work.
}}

\author{\IEEEauthorblockN{Daqing Wu$^{1,2}$, Xiao Luo$^{1,2}$, Zeyu Ma$^{3}$, Chong Chen$^{1,2}$, Minghua Deng$^1$, Jinwen Ma$^{1,*}$}
\IEEEauthorblockA{\textit{$^1$School of Mathematical Sciences, Peking University, 
Beijing, China}\\
\textit{$^2$Damo Academy, Alibaba Group,
Hangzhou, China}\\
\textit{$^3$School of Computer Science and Technology, Harbin Institute of Technology, Shenzhen Graduate School, Shenzhen, China }\\
$\{$wudq,xiaoluo$\}$@pku.edu.cn, zeyu.ma@stu.hit.edu.cn, \\
cheung.cc@alibaba-inc.com, \{dengmh,jwma\}@math.pku.edu.cn}

}

\maketitle

\begin{abstract}
Nowadays, E-commerce is increasingly integrated into our daily lives. Meanwhile, shopping process has also changed incrementally from one behavior (purchase) to multiple behaviors (such as view, carting and purchase). Therefore, utilizing interaction data of auxiliary behavior data draws a lot of attention in the E-commerce recommender systems. However, all existing models ignore two kinds of intrinsic heterogeneity which are helpful to capture the difference of user preferences and the difference of item attributes. First (intra-heterogeneity), each user has multiple social identities with otherness, and these different identities can result in quite different interaction preferences. Second (inter-heterogeneity), each item can transfer an item-specific percentage of score from low-level behavior to high-level behavior for the gradual relationship among multiple behaviors. Thus, the lack of consideration of these heterogeneities damages recommendation rank performance. To model the above heterogeneities, we propose a novel method named intr\underline{A}- and inte\underline{R}-hetero\underline{G}eneity rec\underline{O}mmendation model (ARGO). Specifically, we embed each user into multiple vectors representing the user’s identities, and the maximum of identity scores indicates the interaction preference. Besides, we regard the item-specific transition percentage as trainable transition probability between different behaviors. Extensive experiments on two real-world datasets show that ARGO performs much better than the state-of-the-art in multi-behavior scenarios.

\end{abstract}

\begin{IEEEkeywords}
Recommender Systems, Multi-behavior Recommendation, Collaborative Filtering
\end{IEEEkeywords}

\section{Introduction}
Recently, recommender systems have been widely utilized in online information systems\cite{ricci2011introduction}. Traditional recommender systems only consider the user-item interaction data of one behavior type (i.e., purchase on E-commerce scenarios). Collaborative filtering \cite{sarwar2001item,hu2008collaborative} is the most popular paradigm for building a recommender system, which learns user interest and estimates preference from the collected user behavioral data. However, as the amount and variety of E-commerce data increase, utilizing interaction data of other auxiliary behavior data (such as the click, view, carting and so on) draws much attention. Several studies have utilized these multiple behaviors to provide useful signals of user preferences, which helps to build recommender systems with better performances \cite{zhao2015improving,loni2016bayesian,gao2019neural,chen2020efficient}. 

\begin{figure}
    \centering
    \includegraphics[width=3.3in, keepaspectratio=true]{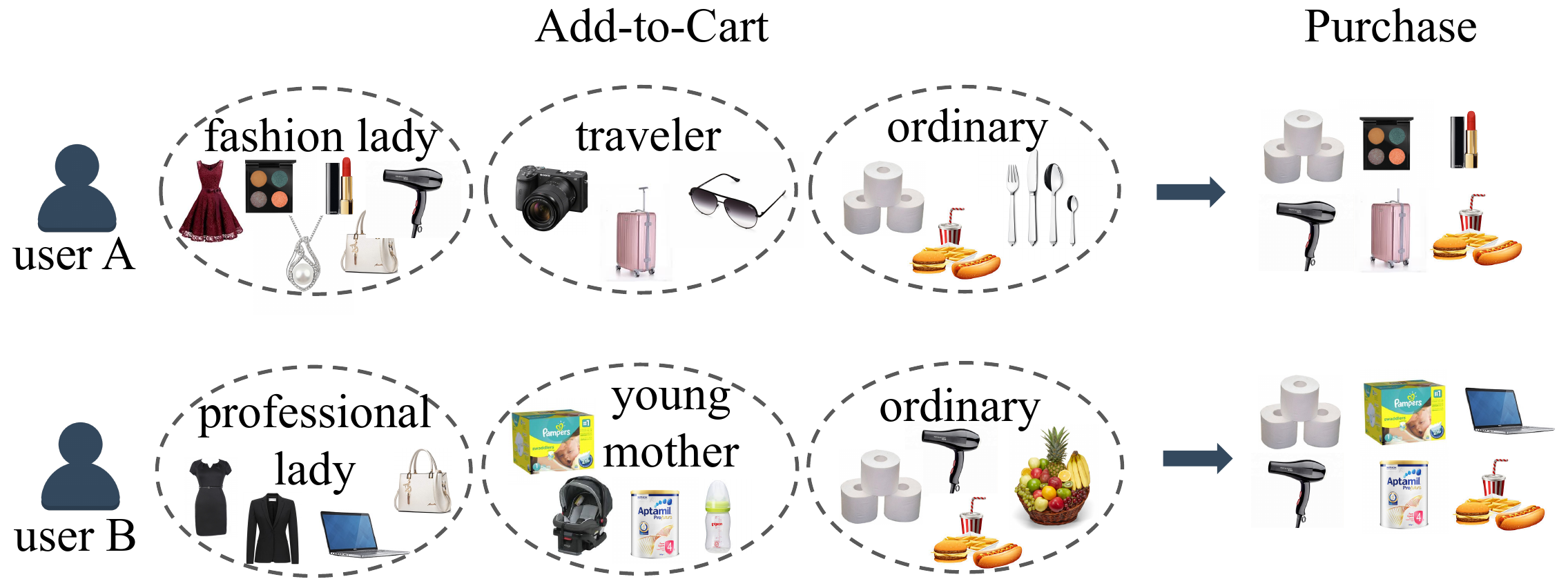}
    \caption{An example of the heterogeneity in E-commerce. Each dashed circle portrays an identity of a user. From the vertical perspective of adding-to-cart behavior, there are several differential identities for user A/B, and each interacted item associates the demand of one identity. From the horizontal view, items such as daily necessities will be purchased with high probability after adding-to-cart, but luxury like jewelry is not in reverse.}
    \label{fig:example}
    \vspace{-4mm}
\end{figure}

Deep learning has achieved much success in the field of computer vision and natural language processing \cite{lecun2015deep, luo2020expectation, luo2020survey}. There are also some attempts on deep learning in the field of recommender systems \cite{hsieh2017collaborative,he2017neural,ebesu2018collaborative,deng2019deepcf,zhang2019star,he2020lightgcn}. Generalized matrix factorization (GMF) \cite{he2017neural} replaces the inner product with embedding layer and fully connected prediction layer, which applies the representation learning on the collaborative filtering with single behavior data. For multi-behavior recommendation, Efficient Heterogeneous Collaborative Filtering (EHCF) \cite{chen2020efficient} is the existing state-of-the-art method. It has shown the superiority than negative sampling (NS) methods, including MC-BPR \cite{loni2016bayesian} and Neural Multi-Task Recommendation (NMTR) \cite{gao2019neural}.

However, in previous works, there is a lack of in-depth consideration of heterogeneity in E-commerce, which is shown in Figure \ref{fig:example}. When faced with the demands of daily work, study and life, each user has multiple social identities. The heterogeneity between different identities reflects the difference of interaction preference of items. Besides, there is a cascading relationship among multiple behaviors in E-commerce scenario (e.g., a user must 
view an item before purchasing it), and the heterogeneity of items determines that each item transfers a specific percentage of score from low-level behavior to high-level behavior. Therefore, from the view of latent space, the previous works which neglect the above two kinds of heterogeneity have two limitations. First, in general, item embedding vectors largely diverge in the latent space, and the user embedding vector should be close to the separated vectors of all positive items, which seems to be unreasonable and thus restricts the performance of deep learning with strong representation ability. Second, the modeling of transferring of previous works merely changes the prediction space from auxiliary behaviors to target behavior for all items, and does not capture item discrepancies between different behaviors, which drops the benefits from the auxiliary behavior data.

To address the above mentioned limitations, we propose a novel model named intr\underline{A}- and inte\underline{R}-hetero\underline{G}eneity rec\underline{O}mmendation model (ARGO) for multi-behavior recommendation task. To cope with the low representation ability of the existing models, we model the intra-heterogeneity by embedding each user into multiple vectors which represent the user's identities, and the vector corresponding to the maximum of identity scores indicates the identity that the interaction preference belongs to. In order to capture the inter-heterogeneity, we consider an ordinal relation among different types of behavior, and relate the prediction of each behavior through an item-specific cascaded manner. To be specific, the prediction of a $(k+1)$-th behavior is obtained from the prediction of the $k$-th behavior as well as learnable item-specific transition probability. Through these designs, our model ARGO effectively captures the user preferences and incorporates the behavior semantics. To summarize, the main contributions of this work are as follows:
\begin{itemize}
    \item In terms of intra-heterogeneity, we propose a reasonable and straightforward neural collaborative filtering model by embedding each user into multiple vectors, which considers the idiosyncrasies of each user.
    \item In terms of inter-heterogeneity, we propose a probability model with trainable transition probabilities, which correlates the prediction of each behavior in an item-specific probabilistic way to capture the ordinal relation among multiple behaviors.
    \item Extensive experiments on two real-world E-commerce datasets show that ARGO outperforms the state-of-the-art models in multi-behavior scenarios by a large margin. 
\end{itemize}

\section{Related Work}

\subsection{Deep Learning-Based Collaborative Filtering}
Since deep neural networks perform well at representation learning, deep learning methods have been widely explored and have shown promising results in various areas such as computer vision \cite{lecun2015deep}. Note that in vanilla matrix factorization, the mapping between the original representation space and the latent space is assumed to be linear, which can not be always guaranteed. To better learn the complex mapping between these two spaces, deep matrix factorization \cite{xue2017deep} utilizes a two pathway neural network architecture to replace the linear embedding operation used in vanilla matrix factorization. NeuMF \cite{he2017neural} replaces the inner product with a neural architecture that can learn an arbitrary function from data, containing the fusion of GMF and Multi-Layer Perceptron (MLP). DeepCF \cite{deng2019deepcf} combines the strengths of representation learning based CF and matching function learning based CF to improve the performance. ConvNCF \cite{he2018outer} uses the outer product instead of the dot product to model user-item interaction patterns and then CNNs are applied over the result of the outer product and could capture the high-order correlations among embedding dimensions.

\subsection{Multi-behavior Recommendation} Multi-behavior based recommendation aims to leverage the behavior data of other types to improve the recommendation performance on the target behavior \cite{loni2016bayesian}. The well-known early model Collective Matrix Factorization model (CMF) \cite{singh2008relational} simultaneously factorizes multiple user-item interactions with sharing item-side embeddings across matrices and extends to leverage multiple user behaviors for recommender systems \cite{zhao2015improving}. NMTR \cite{gao2019neural} accounts for the cascading relationship among different types of behaviors and perform a joint optimization based on the multi-task learning framework, where the optimization on a behavior is treated as a task. EHCF \cite{chen2020efficient} models fine-grained user-item relations as well as efficiently learn model parameters from positive-only data to further improve the performance.

\section{Methodology}
\begin{figure*}[h]
    \centering
    \includegraphics[width=17cm,keepaspectratio=true]{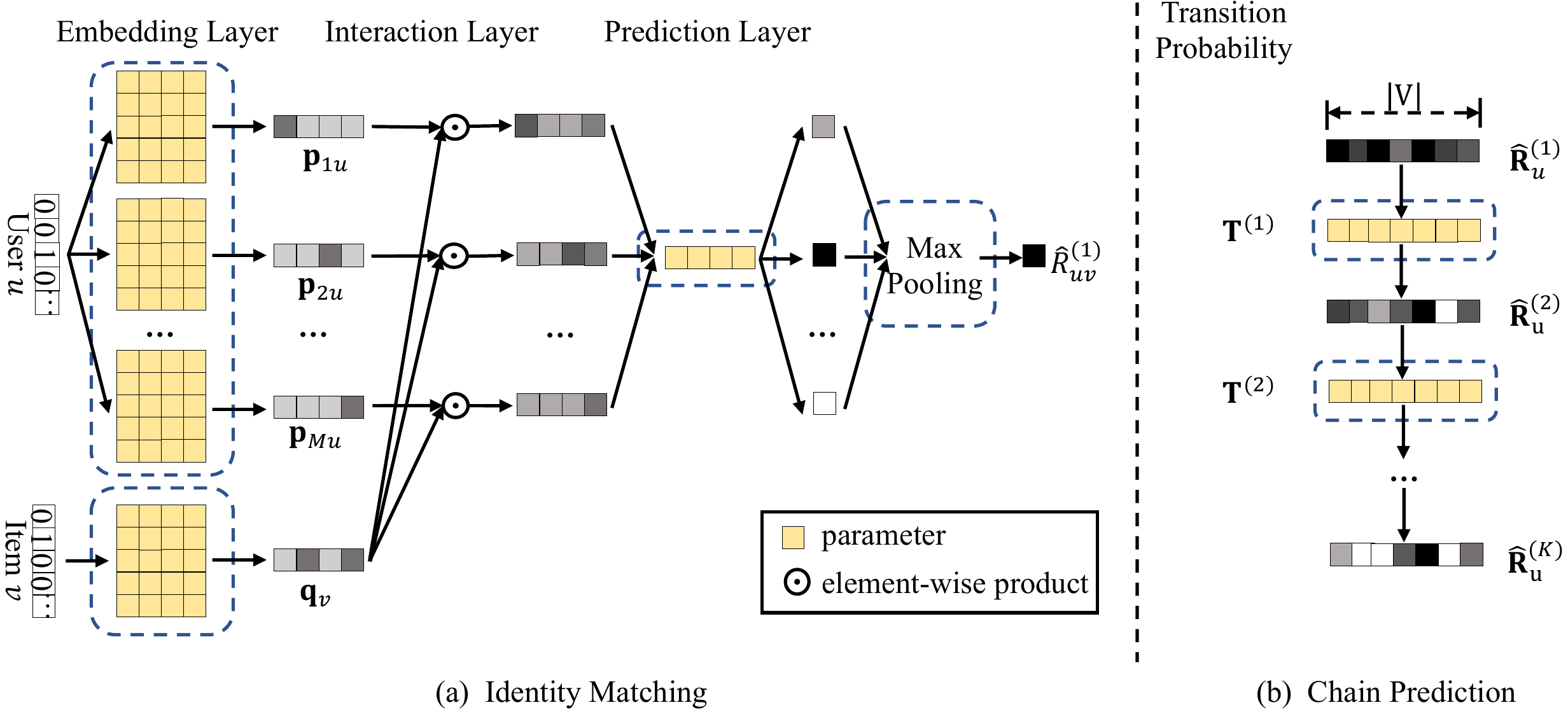}
    \caption{Architecture of ARGO. (a) Given user $u$ and item $v$ as the input, identity matching aims to predict the likelihood that $u$ will perform the first behavior on item $v$, represented as the output of $\hat{R}_{uv}^{(1)}$. (b) On the chain prediction stage, for each user, the likelihood of the $(k+1)$-th behavior is estimated by the product of the likelihood and learnable transition probability of the $k$-th behavior.} 
    \label{fig:framwork}
    \vspace{-4mm}
\end{figure*}
In this section, we first formally define the problem and then feature our ARGO model with two special designs shown in Figure \ref{fig:framwork}:
\begin{itemize}
	\item \textbf{Identity matching (intra-heterogeneity)}. Each user is expressed by $M$ embedding vectors corresponding to $M$ identities, and the maximum of $M$ likelihood is considered as the output for the first behavior.
	\item \textbf{Chain prediction (inter-heterogeneity)}. Low-level behavior will transfer a certain percentage of score to high-level behavior for each item. We correlate the predictions of the ordinal behaviors through chain prediction, modeling by probability model with trainable transition probabilities. 
\end{itemize}

\subsection{Problem Formulation}
Suppose the dataset contains users $U$ and items $V$, $\left\{\mathbf{R}^{(1)}, \mathbf{R}^{(2)}, \ldots, \mathbf{R}^{(K)}\right\}$ denote the user-item interaction matrices of size $|U|\times|V|$ for all $K$ types of behaviors, in which $\mathbf{R}^{(k)}$ with each entry having value 1 or 0 comes from users’ implicit feedbacks of the $k$-th behavior: 
\begin{equation}
    R_{uv}^{(k)}=\left\{
    \begin{array}{ll}
    1, & \text {if user $u$ has observed interaction with item $v$} \\
    0, & \text {otherwise.}
    \end{array}\right.
\end{equation}
Generally, the $K$-th behavior is set to be the target behavior to be optimized in the multi-behavior recommendation task. In the E-commerce dataset, a typical target behavior is the purchase behavior, and other behaviors include viewing, clicking and adding to the cart. The task is estimating the likelihood $\hat{R}_{uv}^{(K)}=f(u,v | \Theta)$ that user $u$ will interact with item $v$ under the target behavior. The items (unobserved under the target behavior) are ranked in the descending order of the likelihood, which provides a Top-$N$ item recommendation list for each user.

\subsection{Identity Matching}
We share the embedding layer of users and items for the modeling of all behavior types. Different from the former methods, each user corresponds to $M$ different $d$-dimensional embedding vectors related to different identities in our model, resulting in $M$ user embedding matrices $\{\mathbf{P}_{m}\}_{m=1}^{M}$. Formally,
\begin{equation}
    \mathbf{p}_{mu} = \mathbf{P}_m^{T} \mathbf{X}_{u}, \quad \mathbf{q}_{v} = \mathbf{Q}^{T} \mathbf{Y}_{v}
\label{equ:embedding}
\end{equation}
where $\mathbf{X}_u$ and $\mathbf{Y}_v$ denote the one-hot feature vectors for user $u$ and item $v$, $\{\mathbf{P}_{m} \in \mathbb{R}^{|U|\times d} \}_{m=1}^{M}$ and $\mathbf{Q}\in \mathbb{R}^{|V| \times d}$ are the user embedding matrices and the item embedding matrix, respectively. Note that embeddings in matrix $\{\mathbf{P}_{m}\}_{m=1}^{M}$ and $\mathbf{Q}$ are served as the initialization feature of users' identities and items, which can be seen as the input feature for each user's identity and item in our framework.

The interaction layer is right behind the embedding layer, which describes the interaction between the user embedding and item embedding. Note that for the user $u$, each $\mathbf{p}_{mu} \in \mathbb{R}^d$ represents the specific identity embedding from the $m$-th view. The task is to find the proper user-embedding from different identities for the item $v$. We first normalize user identity embedding and item embedding:
\begin{equation}
    \begin{aligned}
        \mathbf{\hat{p}}_{mu} = \frac{\mathbf{p}_{mu}}{||\mathbf{p}_{mu}||}, \quad \mathbf{\hat{q}}_{v} = \frac{\mathbf{q}_{v}}{||\mathbf{p}_{v}||}
    \end{aligned}
\label{equ:normalize}
\end{equation}
Then for each user-item instance $(u, v)$ of the lowest-level behavior, the likelihood that user $u$ has the $m$-th interest on item $v$ is estimated by
\begin{equation}
    \begin{aligned}
	    \hat{R}_{muv} = \text{ReLU}(\mathbf{h}^{T}\left(\mathbf{\hat{p}}_{mu} \odot \mathbf{\hat{q}}_{v}\right)) = \text{ReLU}(\sum_{l=1}^{d} h_{l} \hat{p}_{mu, l} \hat{q}_{v, l})
	\end{aligned}
    \label{equ:mapping}
\end{equation}
in which $\odot $ denotes the element-wise product of vectors, $\mathbf{h} \in \mathbb{R}^d$ is prediction vector. Finally, the likelihood that user $u$ has the interest on item $v$ depends on the maximum value likelihood among the $M$ identities. In formulation,
\begin{equation}
	\hat{R}_{uv}^{(1)} = \max_{m\in M}{\hat{R}_{muv}}
\label{equ:maxpooling}
\end{equation}
\subsection{Chain Prediction}
Generally, for a normal purchase behavior of a user, he/she first needs to see a certain product, and then adds it to the shopping cart. Therefore, high-level behaviors (e.g., buy) usually depend on low-level behaviors (e.g., view). To take advantage of this regular pattern, we introduce a transition probability model to model it.

Specifically, for any item $v$, random variable $X^{(k)}$ is described by the interaction between item $v$ and user $u$ of the $k$-th behavior. Denote $p^{(k)}_{v} = P(X^{(k+1)}=1 | X^{(k)}=1)$ and $q^{(k)}_{v} = P(X^{(k+1)}=1 | X^{(k)}=0)$. For any $k < K$, $p^{(k)}_{v}$ and $q^{(k)}_{v}$ can be simply summarized from the observed data by 
\begin{equation}
    \begin{aligned}
        p^{(k)}_{v} &= \frac{\sharp_{u}\{R_{uv}^{(k)}=1, R_{uv}^{(k+1)}=1\}}{\sharp_{u}\{R_{uv}^{(k)}=1\}} \\
        q^{(k)}_{v} &= \frac{\sharp_{u}\{R_{uv}^{(k)}=0, R_{uv}^{(k+1)}=1\}}{\sharp_{u}\{R_{uv}^{(k)}=0\}}
    \end{aligned}
    \label{equ:probability_estimation}
\end{equation}
Note that there is a certain ordinal relation among behavior types in E-commerce, such as a user must view a product (i.e., click the product page) or add a product to cart before he/she purchases it. Therefore, the phenomenon that $R_{uv}^{(k)}=0$ is along with $R_{uv}^{(k+1)} = 0$ always holds for real-world datasets, and we just set $q^{(k)}_{v}=0$ for simplicity. From the interpretation of ranking, the probability $P(X^{(k)}=1)$ can be estimated by the likelihood $\hat{R}_{uv}^{k}$ as $\frac{\hat{R}_{uv}^{k}}{\max_{s}\hat{R}_{us}^{k}}$. Therefore, according to the law of total probability, we formulate $P(X^{(k+1)}=1)$ as 
\begin{equation}
    P(X^{(k+1)}=1) = p^{(k)}_{v} \times \frac{\hat{R}_{uv}^{k}}{\max_{s}\hat{R}_{us}^{k}}
    \label{equ:total_probability_1}
\end{equation}
It's worth noting that the transition probability obtained from the data directly may not be accurate due to the unobserved interaction. For all items, we consider $p^{(k)}_{v}$ as a learnable parameter $t^{(k)}_{v}$ and denote $\mathbf{T}^{(k)} = (t^{(k)}_{1},\cdots, t^{(k)}_{|V|} )$. As a result, the basic interaction prediction is utilized for the lowest-level behavior, and the likelihood of the high-level behavior is estimated level by level:
\begin{equation}
        \hat{\mathbf{R}}_{u}^{(k+1)} = \hat{\mathbf{R}}_{u}^{(k)} \odot \mathbf{T}^{(k)}
    \label{equ:total_probability_2}
\end{equation}
where $\hat{\mathbf{R}}_{u}^{(k)} = (\hat{R}_{u1}^{(k)}, \cdots, \hat{R}_{u|V|}^{(k)})$. With the strong representation of neural networks, the probabilistic transfer vector is refined. Our prediction takes consideration of the effect of behavior-item interaction for each behavior. Note that our model is helpful for predicting the preference of inactive users that have little data on the target behavior because some behaviors (e.g., low-level) are easier to be collected and have a larger volume than the target behavior. Analysis of sparse data will show the effectiveness in experiments.

\subsection{Loss Function and Multi-Task Learning}
Due to the large space of items, the observed interactions are limited, and unobserved interactions are of a large scale in implicit data. To learn model parameters, a weighted regression with squared loss were introduced \cite{hu2008collaborative}:
\begin{equation*}
    L_{r} = \sum_{u \in U}\sum_{v \in V^{+}_{u} \cup V^{-}_{u}} w_{uv}(R_{uv}-\hat{R}_{uv})^2
\end{equation*}
where $V^{+}_{u}$, $V^{-}_{u}$ are positive and negative (missing) items for user $u$, $w_{uv}$ denotes the weight of entry $R_{uv}$. There have been many studies on how to assign proper weights for missing data. Here we adapt a uniform weight $w$ for missing entry \cite{pan2008one}:
\begin{equation}
    L_{r}^{(k)} = \sum_{u \in U}\big(\sum_{v \in {V_{u}^{(k)+}}} {(1-{\hat{R}_{uv}}^{(k)})}^2 + w \sum_{v \in {V_{u}^{(k)-}}} {\hat{R}_{uv}}^{(k)^2}\big)
    \label{equation:l_r}
\end{equation}

Besides, for $m=1,\cdots, M$, $\mathbf{P}_m$ represents the embedding of the $m$-th identity of users. To increase the representation ability of our model, we propose a divergence loss function to penalize the solution when $\mathbf{P}_m$ is almost identical for different $m$. As a consequence, we expect the cosine distance of each identity embedding vector is large enough (i.e., $cos(\mathbf{p}_{mu}, \mathbf{p}_{m'u})\leq 0$). We impose more penalty on the vector pair with closer distance by taking the inflation operation on the cosine similarity. 
\begin{equation}
    L_{d}= \frac{2}{|U|M(M-1)} \sum_{u \in U}\sum_{m\neq m'}(max(0,cos(\mathbf{p}_{mu}, \mathbf{p}_{m'u})))^2
    \label{equation:l_d}
\end{equation}
in which $cos(\cdot)$ means cosine similarity (i.e., the inner product of the two normalized vectors) and the term $2/{|U|M(M-1)}$ is added to calculate the mean. 

Finally, following the paradigm of Multi-Task Learning (MTL) that performs joint training on the models of different but correlated tasks, we obtain the loss function that is minimized as
\begin{equation}
    L = \sum_{k=1}^{K} \lambda_{k} L_{r}^{(k)} +  L_{d} 
    \label{equation:loss}
\end{equation}
where $\lambda_k$ is included to control the influence of the k-th type of behaviors on the joint training. This is a hyper-parameter to be specified for different datasets, since the importance of a behavior type may vary for problems of different domains and scales. We additionally enforce that $\sum_{k=1}^K \lambda_k =1$ to facilitate the tuning of these hyper-parameters.

\section{Experiments}
Now, we conduct experiments on two real-world E-commerce datasets to evaluate our model ARGO, and answer the following four research questions:
\begin{itemize}
	\item \textbf{RQ1}: How does ARGO perform comparing to the baseline models that aim to learn from multi-behavior data?
	\item \textbf{RQ2}: How is the effectiveness of our proposed techniques as well as different types of auxiliary behaviors?  
	\item \textbf{RQ3}: How can ARGO help to alleviate the sparse data problem, i.e., few records on the target behavior?
	\item \textbf{RQ4}: How does the number of identities $M$ affect the final performance of ARGO?
\end{itemize}

\subsection{Datasets}
To evaluate the performance of ARGO, we experiment on two real-world E-commerce datasets: Beibei and Taobao. The statistics of datasets are summarized in Table \ref{tab:dataset}. \textbf{BeiBei}\cite{gao2019neural} This is the dataset obtained from Beibei, the largest infant product E-commerce platform in China. There are 21716 users and 7977 items with three types of behaviors, including purchase, carting, and view collected in this dataset. \textbf{Taobao}\cite{zhu2018learning} This is the dataset obtained from Taobao, the largest E-commerce platform in China. There are 48749 users and 39493 items with three same types of behaviors, comparing to Beibei.

\begin{table}[h]
    \caption{Statistical details of the evaluation datasets.}
    \begin{tabular}{lrrrrrr}
    \hline
    Dataset & \#User & \#Item & \#View & \#Add-to-cart & \#Purchase \\
    \hline
    Beibei & 21,716 & 7,977 & 2,412,586 & 642,622 & 304576\\ 
    Taobao & 48,749 & 39,493 & 1,548,126 & 193,747 & 2259747 \\ 
    \hline
    \end{tabular}
    \label{tab:dataset}
\end{table}

\begin{table*}[h]
    \centering
    \small
    \caption{Top$-N$ Recommendation Performance Comparison on Beibei and Taobao}
    \begin{tabular*}{1\textwidth}{@{\extracolsep{\fill}}c|l|cccccccc@{}}
    \toprule
    & \multirow{2}{*}{Models} & \multicolumn{8}{c}{Beibei\qquad \qquad \qquad } \\
    \cline{3-10} 
    & & HR@10 & HR@50 & HR@100 & HR@200 & NDCG@10 & NDCG@50 & NDCG@100 & NDCG@200 \\
    \hline
    \multirow{3}{*}{One-Behavior} & BPR & 0.0822 & 0.2637 & 0.4048 & 0.5710 & 0.0418 & 0.0757 & 0.0986 & 0.1217 \\
    & ExpoMF & 0.0452 & 0.1465 & 0.2246 & 0.3282 & 0.0227 & 0.0426 & 0.0553 & 0.0723 \\
    & NCF & 0.0441 & 0.1562 & 0.2343 & 0.3533 & 0.0225 & 0.0445 & 0.0584 & 0.0757 \\
    \hline
    \multirow{5}{*}{Multi-Behavior} & CMF & 0.0482 & 0.1582 & 0.2843 & 0.4288 & 0.0251 & 0.0462 & 0.0661 & 0.0852 \\
    & MC-BPR & 0.0504 & 0.1743 & 0.2755 & 0.3862 & 0.0254 & 0.0503 & 0.0653 & 0.0796 \\
    & NMTR & 0.0524 & 0.2047 & 0.3189 & 0.4735 & 0.0285 & 0.0609 & 0.0764 & 0.0968\\
    & EHCF & 0.1523 & 0.3316 & 0.4312 & 0.5460 & 0.0817 & 0.1213 & 0.1374 & 0.1535 \\
    \cline{2-10}
    & \textbf{ARGO} & \textbf{0.1697} & \textbf{0.3613}  & \textbf{0.4436} & \textbf{0.5639} & \textbf{0.0882}  & \textbf{0.1386} &\textbf{0.1420}  & \textbf{0.1632} \\
    \bottomrule
    & \multirow{2}{*}{Models} & \multicolumn{8}{c}{Taobao\qquad \qquad \qquad } \\
    \cline{3-10} 
    &  & HR@10 & HR@50 & HR@100 & HR@200 & NDCG@10 & NDCG@50 & NDCG@100 & NDCG@200 \\
    \hline
    \multirow{3}{*}{One-behavior} & BPR & 0.0376 & 0.0708 & 0.0871 & 0.1035 & 0.0227 & 0.0269 & 0.0305 & 0.0329 \\ 
    & ExpoMF & 0.0386 & 0.0713 & 0.0911 & 0.1068 & 0.0238 & 0.0270 & 0.0302 & 0.0334 \\
    & NCF & 0.0391 & 0.0728 & 0.0897 & 0.1072 & 0.0233 & 0.0281 & 0.0321 & 0.0345 \\
    \hline
    \multirow{5}{*}{Multi-Behavior} & CMF & 0.0483 & 0.0774 & 0.1185 & 0.1563 & 0.0252 & 0.0293 & 0.0357 & 0.0379 \\
    & MC-BPR & 0.0547 & 0.0791 & 0.1264 & 0.1597 & 0.0263 & 0.0297 & 0.0361 & 0.0397 \\
    & NMTR & 0.0585 & 0.0942 & 0.1368 & 0.1868 & 0.0278 & 0.0334 & 0.0394 & 0.0537 \\
    & EHCF & 0.0717 & 0.1618 & 0.2211 & 0.2921 & 0.0403 & 0.0594 & 0.0690 & 0.0789 \\
    \cline{2-10} 
    & \textbf{ARGO} & \textbf{0.0793} & \textbf{0.1745}  & \textbf{0.2388} & \textbf{0.3027} & \textbf{0.0443}  & \textbf{0.0633} &\textbf{0.0771}  & \textbf{0.0833} \\
    \bottomrule
    \end{tabular*}
    \label{tab:main_result}
    \vspace{-4mm}
\end{table*}

\subsection{Baselines}
To demonstrate the effectiveness of ARGO, we compare it with several state-of-the-art methods. The baselines are classified into two categories: one-behavior models that only utilize target behavior records, and multi-behavior models that take all kinds of behavior into consideration. One-Behavior Models: \textbf{BPR}\cite{rendle2009bpr}, a widely used pairwise learning method for item recommendation, which optimizes pairwise loss with the assumption that observed interaction should have higher score than unobserved ones. \textbf{ExpoMF}\cite{liang2016modeling}, a whole-data based MF method which treats all missing interactions as negative and weighs them by item popularity. \textbf{NCF}\cite{he2017neural}, a deep learning method which combines MF with MLP for item ranking. Multi-Behavior Models: \textbf{CMF}\cite{zhao2015improving}, it decomposes the data matrices of multiple behavior types simultaneously. \textbf{MC-BPR}\cite{loni2016bayesian}, it adapts the negative sampling rule in BPR and expands BPR for heterogeneous data. \textbf{NMTR}\cite{gao2019neural}, it combines the advances of NCF modeling and utilizes a joint optimization based on the multi-task learning framework. \textbf{EHCF}\cite{chen2020efficient}, it models fine-grained user-item relations and efficiently learn parameters from positive-only data without negative sampling.

\subsection{Parameter Settings}
Our model is implemented in Pytorch V1.4 and trained on a single NVIDIA GeForce GTX TITAN X GPU. We search for the optimal parameters on validation data and evaluate the model on test data. The embedding size $d$ is fixed to 64 for all models. we use mini-batch Adagrad\cite{duchi2011adaptive} as the optimizer, having train batch size fixed to 256, and set the learning rate to $0.05$. The dropout ratio $\rho$ is set to $0.5$ to prevent over-fitting. Uniform negative entry weight (0.1 for Beibei and 0.01 for Taobao) and the MTL coefficients $\lambda_k$ $(\lambda_1 = 1/6, \lambda_1 = 4/6, \lambda_3 = 1/6)$ are all selected as the same with EHCF \cite{chen2020efficient} for fairness. For the baseline, we set the negative sampling ratio to 4 for sampling-based methods, an empirical value showing good performance. 

\subsection{Evaluation Metrics}
We apply the widely used leave-one-out technique\cite{gao2019neural,chen2020efficient} and two widely used metrics, Hit Ratio (HR) \cite{karypis2001evaluation} and Normalized Discounted Cumulative Gain (NDCG) \cite{jarvelin2000ir} are adopted to evaluate the performance of each model. As a recall-based metric, HR measures whether the testing item is in the Top-$N$ list, while NDCG is sensitive to position, which assigns a higher score to hits at a higher position. It's noticed that for a user, our evaluation protocol ranks all unobserved items in the training set and thus the obtained results are more persuasive than ranking a random sampling subset.

\begin{table*}[h]
    \small
    \centering
    \caption{Ablation Study of Auxiliary Data and Novelties on Taobao}
    \begin{tabular*}{1\textwidth}{@{\extracolsep{\fill}}c|l|cccccccc@{}}
    \toprule
     & & HR@10 & HR@50 & HR@100 & HR@200 & NDCG@10 & NDCG@50 & NDCG@100 & NDCG@200 \\
    \hline
    \multirow{3}{*}{Data} & ARGO-CV & 0.0338 & 0.0730 & 0.0938 & 0.1162 & 0.0194 & 0.0279 & 0.0313 & 0.0344 \\ 
    & ARGO-V & 0.0573 & 0.1393 & 0.1751 & 0.2371 & 0.0310 & 0.0428 & 0.0554 & 0.0670 \\ 
    & ARGO-C & 0.0660 & 0.1691 & 0.2043 & 0.2684 & 0.0361 & 0.0569 & 0.0619 & 0.0737  \\
    \hline
    \multirow{3}{*}{Model} 
    & ARGO-IM & 0.0741 & 0.1629 & 0.2202 & 0.2903 & 0.0393 & 0.0582 & 0.0685 & 0.0775  \\
    & ARGO-CP & 0.0753 & 0.1645 & 0.2282 & 0.2913 & 0.0401 & 0.0607 & 0.0683 & 0.0773 \\
    \cline{2-10}
     &\textbf{ARGO} & \textbf{0.0793} & \textbf{0.1745}  & \textbf{0.2388} & \textbf{0.3027} & \textbf{0.0443}  & \textbf{0.0633} &\textbf{0.0771}  & \textbf{0.0833}  \\
    \bottomrule
    \end{tabular*}
    \label{tab2}
    \label{tab:ablation_result}
    \vspace{-4mm}
\end{table*}

\subsection{Overall Performance Comparison (RQ1)}
\begin{figure}[h]
    \centering
    \includegraphics[width=3.4in]{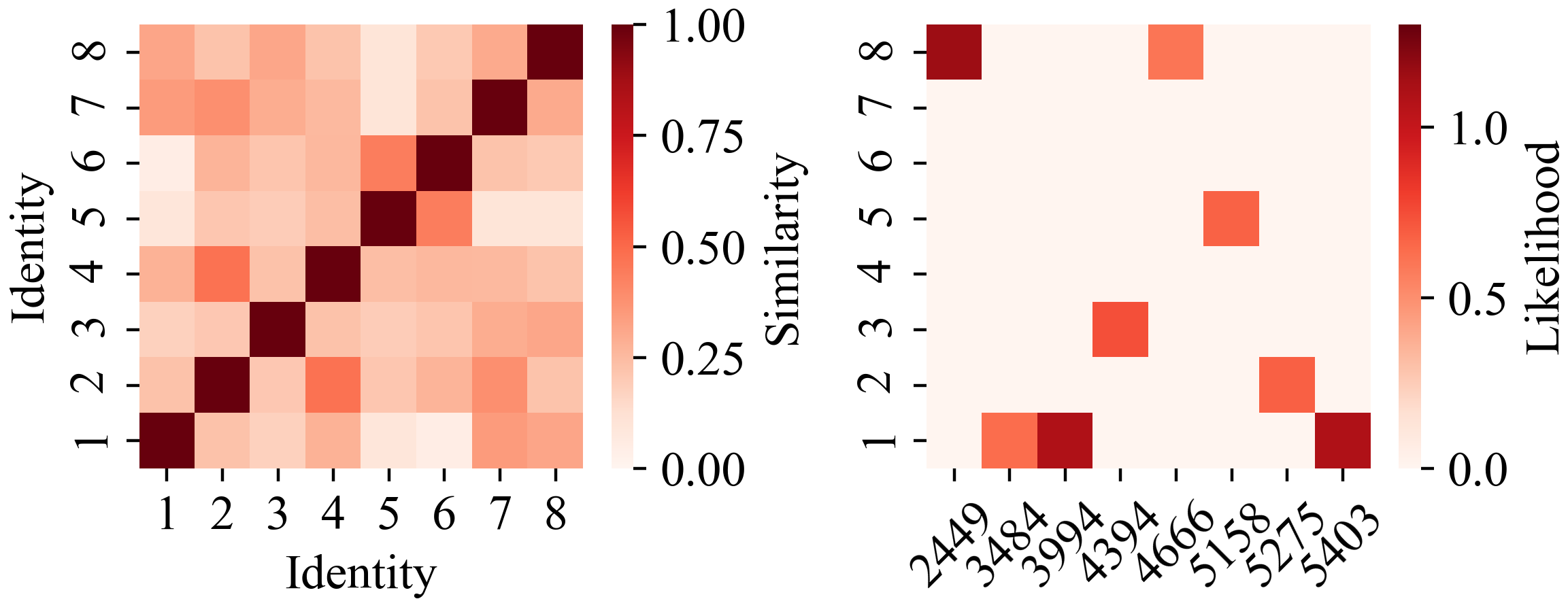}
    \caption{User ID 20137 is selected for instance analysis from Beibei. Left shows the similarities between different identity vectors. Right displays likelihood for identities and positive items.
    \label{fig:interest_similarity}}
\end{figure}

The results of the comparison of different methods on both two datasets are shown in Table \ref{tab:main_result}. We investigate the Top$-N$ performance with $N$ setting to $[10, 50, 100, 200]$. From the results, the following observations can be made:
\begin{itemize}
    \item The methods using multi-behavior feedback generally outperform methods that only make use of purchase behavior, which shows user auxiliary feedbacks improve the models' performances.  The best multi-behavior model among the baselines (EHCF) can outperform the best one-behavior model (NCF) on Beibei by 83.9\% on HR@100 and 135.3\% on NDCG@100, while 83.4\% and 115.0\% on Taobao, which demonstrates the benefits of adding multi-behavior data into the model.
    \item The methods using whole-data-based learning strategies (ARGO, EHCF) generally perform better than other NS methods (NMTR, MC-BPR), which demonstrates that the sampling method leads to a biased result. For example, NMTR fails to capture enough collaborative filtering signals since it adopts the BPR loss function that limits the performance of the model.
    \item We can find that our ARGO obtains the better performance compared to the best baseline EHCF under all metrics. To be specific, the average improvement of our model to EHCF is $5.90\%$ and $7.97\%$ for HR and NDCG on Beibei and $7.52\%$ and $8.45\%$ on Taobao, which justifies the superiority of our model. EHCF uses whole-data based learning strategies but ignores both inter-heterogeneity and intra-heterogeneity while other methods use biased sampling strategies which greatly limit the performances. Our model can fully capture the heterogeneities for target interaction prediction, which helps our model outperforms state-of-the-art multi-behavior recommendation models.
\end{itemize}
\textbf{Case Study}. To further study the impact of identity matching design in ARGO, we select the user (ID 20137) from Beibei, which has interaction with the items "2449, 3484, 3994, 4394, 4666, 5158, 5275" in the training set, and the item "5403" in the test set. From Figure \ref{fig:interest_similarity}, we find that the different identities have relatively large cosine distance, and only one identity embedding has a positive likelihood for each positive interaction. This shows that our identity matching design really captures multiple identities for each user. 

\subsection{Ablation Study (RQ2)}
To understand the effectiveness of two novel  designs and multiple auxiliary behaviors, we conduct experiments with several variants of ARGO. Particularly, we introduce the following model variants: ARGO-CP. Here we replace the chain prediction with three different prediction layers ($h_k$), which are randomly initialized and independent (see \cite{chen2020efficient}). ARGO-IM. Here we only use one embedding vector to represent each user (i.e., $M=1$). ARGO-C. Here we remove the cart behavior information and use view data as the only auxiliary behavior data. ARGO-V. Here we remove the view behavior information and use cart data as the only auxiliary behavior data. ARGO-CV. Here we remove both the view and the cart behavior information and the model only contains the purchase behavior. The results on Taobao are recorded in Table \ref{tab:ablation_result} and the results on Beibei are similar. We summary the following findings:
\begin{itemize}
    \item From the comparison between ARGO and ARGO-CP, we find that the utilization of chain prediction can improve HR consistently. This is reasonable because the chain prediction arranges more positive items in the top-200 list, benefiting from the retention of the likelihood of positive interactions on previous behaviors.
    \item ARGO is consistently superior to ARGO-IM, which is consistent with our above analysis that multiple user embeddings can fit better than single user embedding in the latent space.
    \item ARGO using all types of interaction behaviors consistently outperforms ARGO-V and ARGO-C under all settings, which validates that our model improves purchase forecasting through integrating multi-behavior relations with the MTL framework.
    \item Between the variants that use only one additional behavior type, ARGO-C shows better performance on target prediction in terms of HR@100, which demonstrates the higher importance and effectiveness of utilizing viewing feedbacks for purchasing modeling.
\end{itemize} 

\begin{table*}[h]
    \centering
    \small
    \caption{Performance Comparison of Subset with 5$\thicksim$8 Purchase Records}
    \begin{tabular*}{1\textwidth}{@{\extracolsep{\fill}}l|cccccccc@{}}
    \toprule
    \multirow{2}{*}{Models} & \multicolumn{8}{c}{Beibei\qquad \qquad \qquad  }\\
    \cline{2-9} 
    & HR@10 & HR@50 & HR@100 & HR@200 & NDCG@10 & NDCG@50 & NDCG@100 & NDCG@200 \\
    \hline
    EHCF & 0.1600 & 0.3369 & 0.4424 & 0.5571 & 0.0887 & 0.1274 & 0.1445 & 0.1605 \\
    \textbf{ARGO} & \textbf{0.1701} & \textbf{0.3768}  & \textbf{0.5028}  & \textbf{0.6161} & \textbf{0.1004} & \textbf{0.1455}  & \textbf{0.1656}  & \textbf{0.1815}\\
    \bottomrule
    \multirow{2}{*}{Models} & \multicolumn{8}{c}{Taobao\qquad \qquad \qquad  }\\
    \cline{2-9} 
    & HR@10 & HR@50 & HR@100 & HR@200 & NDCG@10 & NDCG@50 & NDCG@100 & NDCG@200 \\
    \hline
    EHCF & 0.0897 & 0.2060 & 0.2737 & 0.3480 & 0.0492 & 0.0744 & 0.0854 & 0.0958 \\
    \textbf{ARGO} & \textbf{0.0951} & \textbf{0.2228}  & \textbf{0.3060}  & \textbf{0.3677} & \textbf{0.0527} & \textbf{0.0881}  & \textbf{0.1083}  & \textbf{0.1156}\\
    \bottomrule
    \end{tabular*}
    \label{tab:sparse_result}
    \vspace{-4mm}
\end{table*}

\subsection{Effectiveness Analysis on Sparse Data (RQ3)}
Data sparsity is a big challenge for recommender systems based on implicit feedbacks \cite{yin2017mobi}, and multi-behavior recommendation is a typical solution to it. Thus, we study how our proposed ARGO alleviates the problem for those users having few records of the target behavior. Specifically, we collect users with 5$\thicksim$8 purchase records, and there are 6056 and 11846 users on Beibei and Taobao, respectively. Compared to the best baseline EHCF, we conduct experiments of our proposed techniques. The results are shown in Table \ref{tab:sparse_result}. From Table \ref{tab:sparse_result}, we can find HR of ARGO outperforms EHCF by a large margin consistently. Since ARGO models heterogeneous behavior relations in a reasonable way, it can achieve good performance for users with sparse interactions.

\begin{figure}[h]
\centering
\includegraphics[width=3.4in]{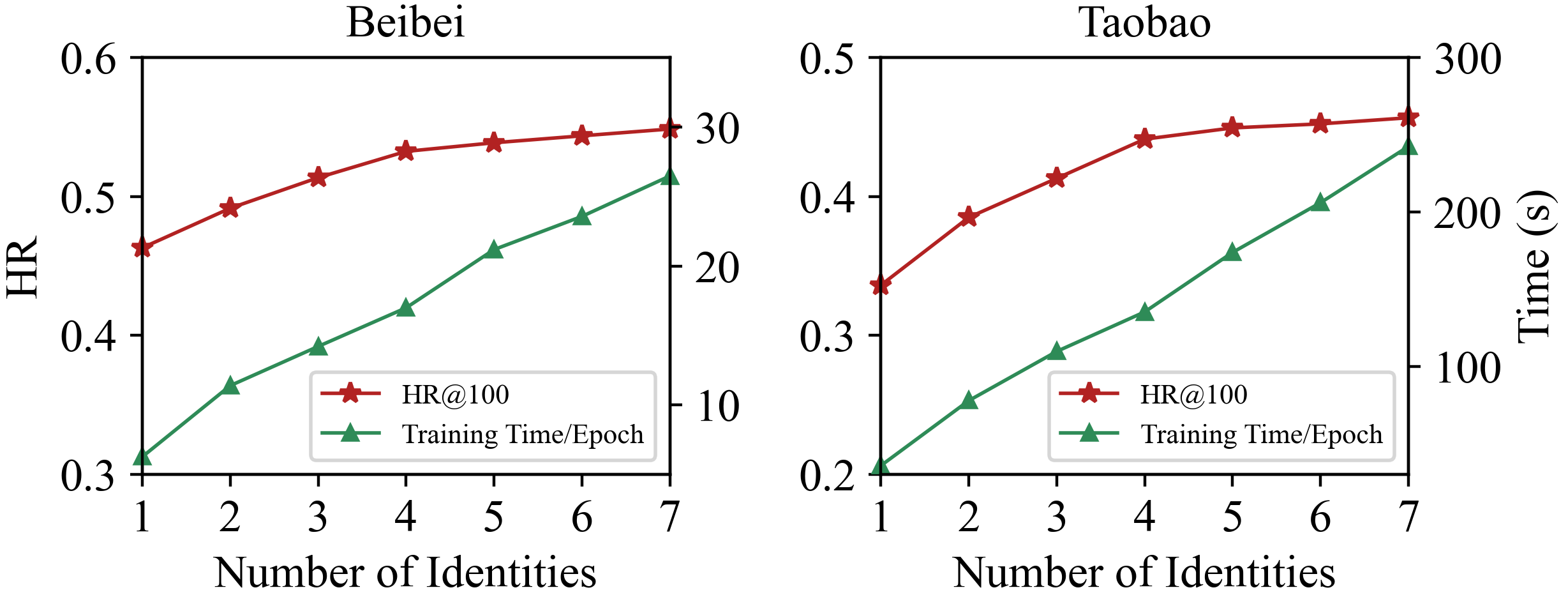}
\caption{HR@100 Performance and runtime of ARGO with different number of identities on the Beibei and Taobao. 
\label{fig:num_interest}}
\end{figure}

\subsection{Parameter Sensitivity (RQ4)}
To understand how hyper-parameters influence the performance of our model, we test the impact of the number of identities $M$, which is related to the representation and generalization ability. We tune $M$ from 1 to 7. The results of HR@100 and running time are plotted in Figure \ref{fig:num_interest}. From the figure, we find that larger $M$ leads to a better performance according to the fact that a large $M$ implies an over-parameterized model which helps the training process with SGD\cite{soltanolkotabi2018theoretical}. However, the training time increases linearly with $M$ and the increasement of HR@100 is scant when $M$ is larger than 4. Therefore, we set $M$ to 4 as default from the trade-off between performance and computational efficiency. 

\section{Conclusion}
In this paper, we propose ARGO for recommendation with heterogeneous user feedback. ARGO has two key characteristics: First, it represents each user with multiple vectors encoding the user identities. Second, the prediction of each behavior is correlated by a learnable chain prediction transition probability model. Extensive experiments on two real-world datasets show that ARGO outperforms the state-of-the-art recommendation models. This work further explores the more complex relationship of different behaviors and opens up a new avenue of research by introducing a probabilistic model for the multi-behavior recommendation. Future work includes exploring our model in complex situations such as cold start problem and knowledge-based recommendation problem. We will also try to extend our method to make it applicable in other recommendation tasks such as sequence-based recommendation.

\section{Acknowledgements}
This work was partially supported by the National Key Research and Development Program of China under grant 2018AAA0100205.

\bibliographystyle{IEEEtran}
\bibliography{main}


\end{document}